\documentclass[prd, showpacs,amsmath,aps,onecolumn,superscriptaddress]{revtex4-1}
\usepackage{amsfonts}
\usepackage{graphicx,color}
\usepackage{times}

\begin{document}
\title{Warm and Cold Dark Matter in Bouncing Universe}
\author{Changhong Li}
\email{changhongli@ynu.edu.cn}
\affiliation{Department of Astronomy,  Key Laboratory of Astroparticle Physics of Yunnan Province, School of Physics and Astronomy, \\ Yunnan University, No.2 Cuihu North Road, Kunming, China 650091}

\begin{abstract}
Being free from the initial singularity, bouncing cosmology serves as a promising alternative to inflation. However, the entropy production during the post-bounce phase is still unclear. In this work, we use a newly unveiled DM candidate (EQFIDM), which can directly {\bf freeze in} thermal equilibrium shortly after bounce, to trace the entropy-producing decay process of residual bouncing field (RBF). Specifically, we present a model-independent formalism to obtain all nine possible types of the entropy-producing decay of RBF within a generic bouncing universe, and show that, due to the entropy productions in different types, the particle mass of this new candidate can range from sub-keV scale to super-TeV scale. With current Lyman-$\alpha$ forest observation on linear matter spectrum, we find that, although four types have been excluded, the rest five types, which suggest warm and cold EQFIDM candidates with effective entropy productions, are accommodated with current observation. Then we illustrate how the nature of RBF can be constrained by current observation and investigate the potential of EQFIDM on alleviating small-scale crisis.

\end{abstract}

\date{\today}

\maketitle

\section{Introduction} 

As the leading scenario of early Universe, inflation not only solves the horizon and flatness problems but also  can generate a nearly scale-invariant curvature perturbation that agrees well with current observations in CMB \cite{Guth:1980zm, Mukhanov:1990me, Komatsu:2010fb, Ade:2015xua}. However, it inevitably suffers from the initial singularity problem \cite{Borde:1993xh}. Due to this concern, bouncing cosmology, which addresses this issue with a non-zero minimal size of universe, has been extensively studied such as in \cite{Khoury:2001wf, Gasperini:2002bn, Creminelli:2006xe, Peter:2006hx, Cai:2007qw, Cai:2008qw,  Li:2011nj, Cai:2011tc, Easson:2011zy, Bhattacharya:2013ut, Qiu:2015nha, Cai:2016hea, Barrow:2017yqt, deHaro:2017yll, Ijjas:2018qbo, Boruah:2018pvq, Nojiri:2019yzg}. As a promising alternative to inflation, it can also resolve the horizon and flatness problems and generate a nearly scale-invariant curvature spectrum \cite{Novello:2008ra, Brandenberger:2016vhg, Nojiri:2017ncd}. Moreover, a duality have been established between inflation and bounce, which indicates that, for each inflation model, the corresponding bouncing universe model can generate a curvature perturbation with the same spectral index $n_s$ \cite{Wands:1998yp, Finelli:2001sr, Boyle:2004gv, Li:2013bha}. Therefore, to concretely distinguish inflation and bounce, many efforts have been made besides CMB spectrum \cite{Cai:2009rd, Gao:2014eaa, Ben-Dayan:2016iks, Celani:2016cwm,Clifton:2017hvg, Ni:2017jxw, Chen:2018cgg, Zhang:2019tct, Elizalde:2020zcb, Frion:2020bxc}. 

One promising proposal is the ``big-bounce genesis'' \cite{Li:2014era}. In which, thermally produced bosonic dark matter (DM) candidate is considered as a novel probe for bouncing universe. As it shows, beside WIMP candidate \cite{Lee:1977ua, Steigman:1984ac}, a non-equilibrium DM candidate (NEQDM), which is belong to freeze-in DM (FIDM) \cite{Dodelson:1993je, Chung:1998ua, Shi:1998km, Feng:2003xh, Hall:2009bx, Klasen:2013ypa, Feldstein:2013uha, Baer:2014eja}, can also be produced in a generic bouncing universe. Specifically, due to the tiny cross-section, the abundance of NEQDM never attain thermal equilibrium, so its relic abundance, which encodes the primordial information of early universe, can be used to distinguish inflation and bouncing universe. 

Encouraged by this finding, the big-bounce genesis has also been applied to fermionic DM candidates \cite{Cheung:2014nxi}. In which, WIMP and NEQDM avenues are re-obtained for fermionic candidates. However, thermal decoupling condition of fermionic candidate in bouncing universe has not been quantitatively analyzed, which results in an omission about the new DM candidate unveiled in this paper. More specifically, as demonstrated in \cite{Li:2014era}, the coincidence of equilibrium condition and freeze-out condition for bosonic DM candidates only allows two different thermally-producing avenues,  WIMP and NEQDM. But for other candidates, such coincidence may not exist. For instance, by taking a DM candidate with a temperature-independent thermally averaged cross-section, a novel avenue has been unveiled between WIMP and NEQDM \cite{Li:2014cba}. Inspired by this, we are motivated to search such avenue for fermionic DM candidates in bouncing universe.    

In this paper, by carefully examining thermal decoupling condition, we unveil this novel avenue for fermionic DM candidate within a generic bouncing universe. Through it, the thermal equilibrium freeze-in DM (EQFIDM) candidate can be produced to attain thermal equilibrium and directly freeze in thermal equilibrium shortly after the bounce. The cross-section of EQFIDM can range from the upper bound of NEQDM to the lower bound of WIMP, and its energy density fraction is proportional to the particle mass, $\Omega_\chi\propto \langle \sigma v\rangle^0 m_\chi$, which is different from the well-known predictions of WIMP, $\Omega_\chi^{WIMP}\propto \langle \sigma v\rangle^{-1} m_\chi^{0}$ \cite{Lee:1977ua}, and of NEQDM, $\Omega_\chi^{NEQDM} \propto \langle \sigma v\rangle m_\chi^2$ \cite{Li:2014era}. 

As we will show, for a very tiny cross-section, EQFIDM can freeze in thermal equilibrium at a very early time, so that the post-freezing-in entropy production may be not negligible and eventually affect the relic abundance of EQFIDM. Therefore, EQFIDM can be employed to trace the entropy production during the post-bounce phase. In particular, we present a model-independent formalism to obtain all nine possible types of entropy-producing decay processes of residual bouncing field (RBF) within a generic bouncing universe. Our result indicates that, in these types with different strength of entropy production, the particle mass of EQFIDM can range from sub-keV scale to super-TeV scale. 

With current Lyman-$\alpha$ forest observation on linear matter spectrum \cite{Markovic:2013iza}, we find that, although four types have been excluded for their tiny entropy production, the rest five types, which suggest warm and cold EQFIDM candidates with effective entropy production, are accommodated with current observation. Especially, we find that, if RBF can be re-dominant after freeze-in for a while, its entropy production is solely determined by the fundament nature of RBF rather than the energy density ratio at freezing-in. Then we use it to illustrate how the nature of RBF can be constrained by current Lyman-$\alpha$ forest observation. At the end, we also investigate the potential of EQFIDM on alleviating the well-known small-scale crisis \cite{Weinberg:2013aya}.  

The reminder of this paper consists of four sections. In section II, we analytically solve the Boltzmann equation within a generic bouncing universe and demonstrate the existence of EQFIDM avenue. In section III, we present a model-independent formalism to  obtain all nine possible types of entropy-producing decay of RBF. In section IV, we use current Lyman-$\alpha$ forest observation to constrain these types and investigate the potential of EQFIDM on alleviating SSC. We conclude in the last section.

\section{Thermal equilibrium freeze-in dark matter candidate in Bouncing Universe} 
For thermally produced DM candidate, the evolution of its abundance is govern by Boltzmann equation,
\begin{equation} \label{eq:blze}
\frac{d (n_\chi a^3)}{a^3dt}=\widetilde{\langle\sigma v\rangle}\left[\left(n_\chi^{eq}\right)^2-n_\chi^2\right]
\end{equation}
where $a$ is the scale factor, $n_\chi$ and $\widetilde{\langle\sigma v\rangle}$ are, respectively, the number density and thermally averaged cross-section of DM particles, and the superscript $~^{eq}$ denotes thermal equilibrium.  

\begin{figure}[htp!]
\centering
\includegraphics[width=0.8\textwidth]{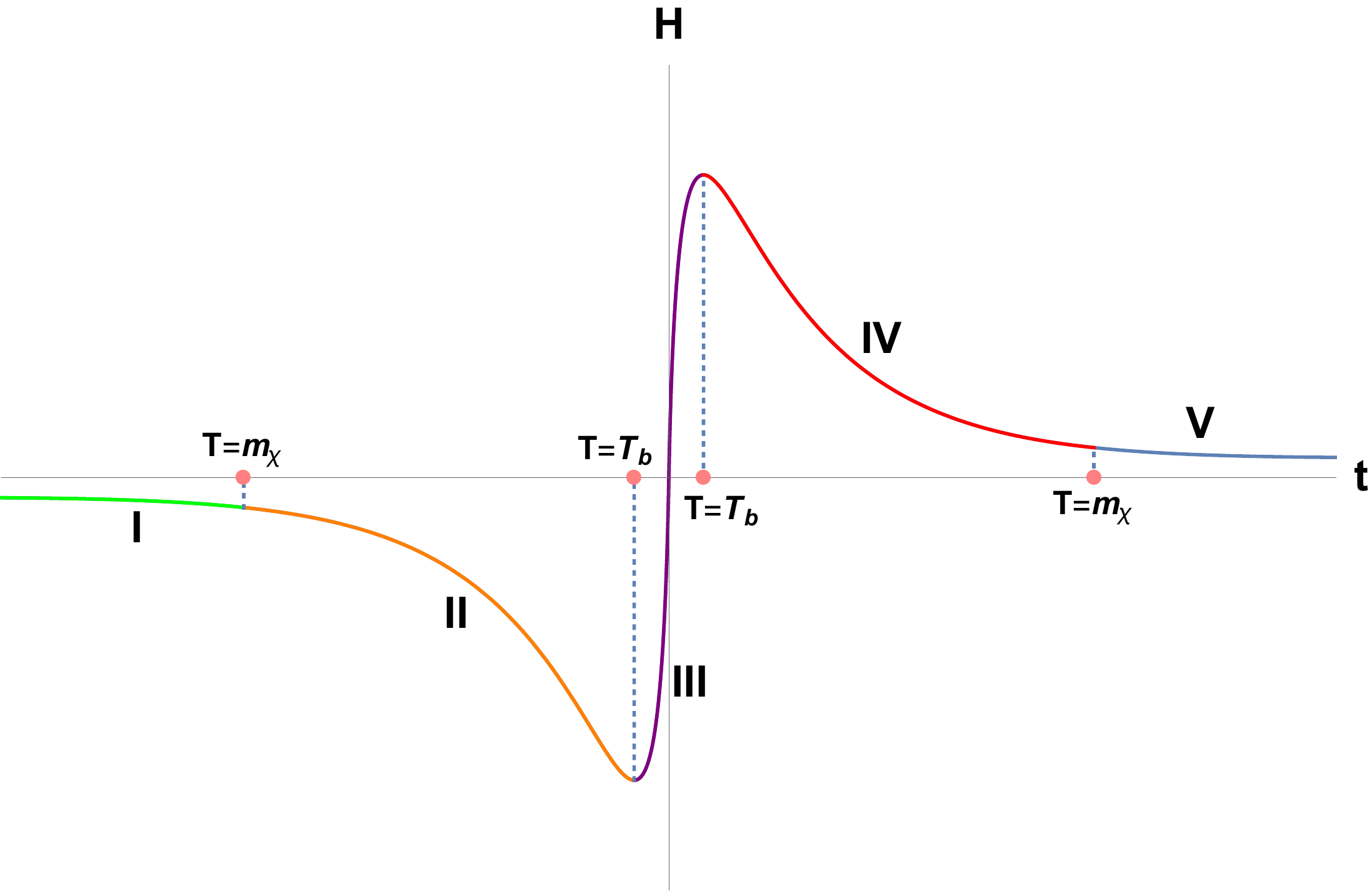}
\caption{Five stages of a generic bouncing universe with respect to DM evolution. From left to right, there are I. pre-production phase (green), II.  contracting phase (orange), III. bouncing region (purple), IV.  expanding phase (red), and V. freezing-out/in phase (blue). }
\label{fig:Bounce.pdf}
\end{figure}

To analytically solve Eq.(\ref{eq:blze}), we divide a generic bouncing universe into five stages as illustrated in Fig.\ref{fig:Bounce.pdf}.
\begin{itemize}
\item {\bf I. The Pre-production Phase} ($T< m_\chi$ \& $H<0$). From the initial low temperature Bunch-Davies vacuum ($T\simeq 0$),  the universe collapses to get hotter and eventually becomes radiation-dominated \cite{Cai:2011ci}. To ensure no effective DM production taking place in this phase, we take $T\simeq m_\chi$ to denote the end of this phase. 
%%%%%%
\item {\bf II. The Contracting Phase} ($m_\chi \le T\le T_b $ \& $H<0$). During this phase, dominated by cosmic bath, the universe continues to contract. With $T\ge m_\chi$, DM particles are effectively produced (WIMP and EQFIDM immediately attains thermal equilibrium with large $\langle \sigma v\rangle$ and NEQDM continues to be produced with small $\langle \sigma v\rangle$). With further contraction, the background temperature of the universe continues to increase until it finally attains the critical value, $T=T_b\gg m_\chi$, at the end of this phase. After that, the non-standard-physics-inspired bouncing field becomes dominant and gives rise to bounce. .

%%%%%%
\item {\bf III. The Bounce Phase} ($ T\ge T_b$ \& $-|H(T_b)|<H<|H(T_b)|$).  In this phase, the dominant bouncing field drives the universe bouncing from contraction ($^-$) to expansion ($^+$) . Specifically, after the Universe contracts to its minimal size, it starts to expand. When the background temperature reaches $T=T_b$ in expansion, the cosmic bath becomes dominant again and this phase ends. Not that, for a robust bounce, the bouncing field is restored  and no entropy-producing decay takes place during this phase \cite{Cai:2011ci}, which leads to the matching condition, $Y^{+}(T_b)=Y^{-}(T_b)$, for WIMP and EQFIDM \footnote{For NEQDM, it continues to be produced during the bounce as it has not attained thermal equilibrium, so that its matching condition should be $Y^{+}(T_{max})=Y^{-}(T_{max})$, where $T_{max}$ is the background temperature when the universe attains its minimal size $a_{min}$. For an efficient bounce, we have $a_{min}\sim a(T_b)$ and $T_b\sim T_{max}$, which can also lead to $Y^{+}(T_b)=Y^{-}(T_b)$. Only if the bounce is not efficient, $T_b\ll T_{max}$, such correction should be taken into account for NEQDM. In this paper, as we mainly focus on EQFIDM, we will use $Y^{+}(T_b)=Y^{-}(T_b)$ for all three candidates in our calculation for simplicity.}, where $Y\equiv n_\chi/T^3$ is DM abundance.

 %%%%%%
\item {\bf IV.  The Expanding Phase} ($T_b\ge T\ge m_\chi$ \& $H> 0$). After the bounce phase, the universe becomes radiation-dominated and continues to expand. With the falling of background temperature, EQFIDM with very tiny cross-section  relativistically freezes in thermal equilibrium at a very early time. Meanwhile, since the beginning of this phase, the residual part of bouncing field also starts to decay and heats cosmic bath mildly. If RBF is depleted after EQFIDM freezes in, a part of its entropy-producing decay will affect the relic abundance of EQFIDM, so we can use EQFIDM to trace RBF. This phase ends at $T=m_\chi$, at which the non-relativistic thermal decoupling takes place. 
 
\item and {\bf V. The Freezing-out/in Phase} ($m_\chi>T$ \& $H>0$). In this phase, with the non-relativistic thermal decoupling, WIMP starts to freeze out and NEQDM starts to freeze in. Note that,  such non-relativistic thermal decoupling can not affect EQFIDM as EQFIDM has already relativistically frozen in thermal equilibrium earlier.
\end{itemize}

According to above analysis, DM particles are mainly produced in Phase II and Phase IV. In these two phases, we have two scaling relations,  $H=\pm H_my^2$ for Hubble parameter and  $\widetilde{\langle\sigma v\rangle}=\langle\sigma v\rangle y^{-n}$ for thermally averaged cross-section, which can significantly simplify Eq.(\ref{eq:blze}),
\begin{equation}\label{eq:abrd}
\frac{d Y}{dy}=\mp \kappa y^{-n-2}(1-\pi^4 g_\chi^{-2} Y^2), 
\end{equation}
where  $y\equiv m_\chi/T$, $\kappa\equiv g_\chi^2\langle\sigma v\rangle m_\chi^3\pi^{-4}H_m^{-1}$ is a newly introduced dimensionless constant with $g_\chi$ being the degree of freedom of DM particle, $H_m$ and $\langle\sigma v\rangle$ are constant, and $n =\pm 2$, respectively, corresponds to fermion and boson \cite{Gondolo:1990dk}. Note that, for bosonic candidate, $n=-2$ leads to a coincidence for the equilibrium condition and freeze-out condition, so there is no other candidate between WIMP and NEQDM. This is the reason why EQFIDM has not been found in the original literature of big-bounce genesis \cite{Li:2014era}. As we will show in this paper, for fermionic candidate with $n=2$, there is no such coincidence , so that, between WIMP and NEQDM, EQFIDM avenue exists for fermionic DM candidate in bouncing universe.        

By solving Eq.(\ref{eq:abrd}) with the initial condition $Y_-(y=1)=0$ and using the matching condition $Y_-(T_b)=Y_+(T_b)$, we can obtain DM abundance at the end of Phase IV, 
\begin{equation}\label{eq:sddmp}
 Y_+(y=1)=\bar{Y}_\chi^{eq}\tanh\left[\frac{\pi^2}{g_\chi}\frac{2\kappa}{3}\left(\frac{1}{y_b^{3}}-1\right)\right],
\end{equation}
where $\bar{Y}_\chi^{eq}=g_\chi\pi^{-2}$ is the thermal equilibrium abundance in the radiation-dominated phase. The factor $2$ in front of $\kappa$ reflects that non-equilibrium DM particles can be produced in both of the contracting and expanding phases. 

By checking whether or not DM abundance attains thermal equilibrium at $y=1$, we can divide Eq.(\ref{eq:sddmp})  into two categories, 
\begin{equation}  \label{eq:Yplus2}
Y_+=
\left\{  
  \begin{array} {lr}
 {\displaystyle \bar{Y}_\chi^{eq},   \qquad \tilde{\kappa}\gg 1}, ~\text{\bf Equilibrium} 
 \\ 
  {\displaystyle \bar{Y}_\chi^{eq}\tilde{\kappa}, \quad~ \tilde{\kappa}\ll 1}, ~~\text{\bf Non-equilibrium}   
  \\
\end{array}     
\right. .
\end{equation}
 where $\tilde{\kappa}\equiv \frac{\pi^2}{g_\chi}\frac{2\kappa}{3}y_b^{-3}\propto \langle \sigma v\rangle  T_b^3$. This result indicates that, for a given value of $T_b$, only large cross-section DM candidates can attain thermal equilibrium.  

With the falling of background temperature, the non-relativistic thermal decoupling takes place at $y=1$. Using $\widetilde{\langle\sigma v\rangle}\rightarrow \langle\sigma v\rangle$ and $Y^{eq}\rightarrow 0$ in this transition, Eq.(\ref{eq:blze}) is simplified as 
\begin{equation} \label{eq:sbefo}
\frac{dY}{dy}=-\kappa  \frac{\pi^4}{ g_\chi^{2}} \frac{Y^2}{y^2}~,
\end{equation}
which solution takes
\begin{equation}
Y(y\rightarrow \infty)= \left(\frac{3}{2}\frac{\pi^2}{g_\chi}y_b^3\tilde{\kappa}+Y_+^{-1}\right)^{-1}.
\end{equation}
Again, this complete solution can be divided into two categories,
\begin{equation}  \label{eq:Yplus3}
Y_f=
\left\{  
  \begin{array} {lr}
 {\displaystyle \tilde{\kappa}^{-1} \frac{2g_\chi}{ 3\pi^{2}y_b^{3}},  \quad~ \tilde{\kappa}\gg  Y_+^{-1} \frac{2g_\chi}{ 3\pi^{2}y_b^{3}}}, \text{\bf ~Freezing-out} 
 \\ 
  {\displaystyle  Y_+,   \qquad \qquad ~~ \tilde{\kappa}\ll Y_+^{-1} \frac{2g_\chi}{ 3\pi^{2}y_b^{3}}}, ~~\text{\bf Freezing-in}   
  \\
\end{array}     
\right. .
\end{equation}
 
Notably, in contrast to bosonic DM candidates \cite{Li:2014era}, Eq.(\ref{eq:Yplus2}) and Eq.(\ref{eq:Yplus3}) are not coincided for fermionic DM candidates. It thus implies that, between WIMP and NEQDM ($1\ll \tilde{\kappa}\ll \frac{2}{3}y_b^{-3}$), there is new  fermionic candidate (EQFIDM), which undergoes Equilibrium Production and Freezing-in processes. More specifically, as listed in Table~\ref{tab:cetpdm}, WIMP has a very large cross-section, so that it can attain thermal equilibrium and eventually freeze out. NEQDM has a small cross-section, so that it is not able to attain thermal equilibrium and consequently can not freeze out. The novelty for EQFIDM is that its cross-section is large enough to attain thermal equilibrium but too small to freeze out, so it undergoes the intermediate avenue between WIMP and NEQDM. To compare them, in Table~\ref{tab:cetpdm}, we have used the characteristic relation of relic energy density fraction $\Omega_\chi\propto m_\chi Y_f$ to present our result.

\begin{table*}[ht!]  
\vspace{-0.5cm}
\caption{\label{tab:cetpdm}Three avenues in the thermal DM scenario}
\begin{center}
\begin{tabular}{|c|c|c|}
\hline
& Equilibrium Production ($\tilde{\kappa}\gg 1$) & Non-equilibrium Production ($\tilde{\kappa}\ll 1$) \\
\hline
Freezing-out & ${\mathcal{A}}$~:~ ${\text{WIMP candidate}}~~~~~~~$& --- \\
$\tilde{\kappa}\gg Y_+^{-1} \frac{2g_\chi}{ 3\pi^{2}y_b^{3}}$& $\Omega_\chi \propto \langle \sigma v\rangle^{-1} m_\chi^0$&\\
\hline
Freezing-in & ${\mathcal{C}}$~:~ $\text{EQFIDM candidate}$ &${\mathcal{B}}$~:~ $\text{NEQDM candiate}$\\
$ \tilde{\kappa}\ll Y_+^{-1} \frac{2g_\chi}{ 3\pi^{2}y_b^{3}}$& $\Omega_\chi \propto \langle\sigma v\rangle^0 m_\chi$&$\Omega_\chi \propto \langle \sigma v\rangle m_\chi^2$\\
\hline
\end{tabular}
\end{center}
\vspace{-0.5cm}
\end{table*}

For EQFIDM, by inserting the explicit prefactor into its characteristic relation, we obtain 
\begin{equation}  \label{eq:ommcnwdm}
\Omega_\chi=\frac{32.4}{h^2}\left(\frac{g_\chi}{2}\right)\left(\frac{a_{fi} T_{fi}}{a_0 T_0}\right)^3\left(\frac{m_\chi}{1~\text{keV}}\right)~,
\end{equation} 
where the subscripts $_{fi}$ and $_0$ label freezing-in and today respectively. And by using the condition, $1\ll \tilde{\kappa}\ll \frac{2}{3}y_b^{-3}$, we obtain 
\begin{equation}\label{eq:raofcrse}
\frac{3}{2}\frac{\pi^2}{g_\chi}\frac{H_m}{T_b^3}\ll \langle\sigma v\rangle \ll \frac{\pi^2}{g_\chi}\frac{H_m}{m_\chi^3}~,
\end{equation}
which indicates that its cross-section can range from the upper bound of NEQDM to the lower bound of WIMP.

According to Eq.(\ref{eq:ommcnwdm}), $\Omega_\chi$ is not only determined by $m_\chi$ but also by the factor $\left(a_{fi} T_{fi}/a_0 T_0\right)^3$. Therefore, EQFIDM candidates with different $\langle\sigma v\rangle$ may have different $\Omega_\chi$.  The reason is that, the value of $\left(a_{fi} T_{fi}/a_0 T_0\right)^3$ depends on the magnitude of the post-freezing-in entropy production. As EQFIDM candidates with different $\langle\sigma v\rangle$  freezes in thermal equilibrium at different time, they may have  different $\left(a_{fi} T_{fi}/a_0 T_0\right)^3$ and result in different $\Omega_\chi$ . For instance, in the large $\langle\sigma v\rangle$ limit, $\langle\sigma v\rangle \rightarrow \frac{\pi^2}{g_\chi}\frac{H_m}{m_\chi^3}$, EQFIDM can track thermal equilibrium for a very long time and finally freeze in thermal equilibrium at very late time, $T_{fi}\simeq m_\chi$, so only very little entropy can be produced after freezing-in and $\left(a_{fi} T_{fi}/a_0 T_0\right)^3\simeq 1$. On the other hand, in the small $\langle\sigma v\rangle$ limit, $\langle\sigma v\rangle \rightarrow \frac{3}{2}\frac{\pi^2}{g_\chi}\frac{H_m}{T_b^3}$, EQFIDM  tracks thermal equilibrium only for a very short time and then freezes in thermal equilibrium at a much earlier time, $T_{fi}\simeq T_b$. In this case, the post-freezing-in entropy production could be very large and $\left(a_{fi} T_{fi}/a_0 T_0\right)^3\ll 1$. 

In this paper, we mainly consider two entropy-producing processes, which are, respectively, from Standard Model (SM) particles annihilations and RBF decay. Without loss of generality, we can take $T_r\sim \text{100}$MeV to separate the two entropy-producing processes, 
\begin{equation}
\left(\frac{a_{fi} T_{fi}}{a_0 T_0}\right)^3= \left(\frac{a_{fi} T_{fi}}{a_r T_r}\right)^3\left(\frac{a_r T_r}{a_0 T_0}\right)^3.
\end{equation}
By doing so, we can categorize EQFIDM candidates into three categorises,
\begin{itemize}
\item Large $\langle\sigma v\rangle$. For large cross-section ($T_{fi}\ll T_r$), EQFIDM will freeze in thermal equilibrium at very late time, so no entropy is produced after freezing-in, $ (a_{fi} T_{fi}/ a_0 T_0)^3\simeq 1$. By substituting it into Eq.(\ref{eq:ommcnwdm}), we obtain
\begin{equation}\label{eq:mcforlarcrse}
m_\chi=0.004~\text{keV}~,
\end{equation}
where we have used $\Omega_\chi\simeq 0.26$ and $g_\chi=2$. Obviously, such hot DM candidate is not favored by current astrophysical observations \cite{Dodelson:2003ft}. 

\item Middle $\langle\sigma v\rangle$. For middle cross-section ($T_{fi}\sim T_r$), EQFIDM will freeze in thermal equilibrium after the entropy-producing decay of RBF but before SM particles annihilations. Therefore, we have $(a_{fi} T_{fi}/ a_r T_r)^3=1$ and $(a_r T_r/ a_0 T_0)^3 \simeq 1/30$ \cite{Dodelson:2003ft}, which gives
\begin{equation}  \label{eq:mcnanwdm}
m_\chi=(\Omega_\chi/2.2)\text{keV}~.
\end{equation}
Note that, this prediction is degenerate to that in inflation. The reason is that, without entropy-producing decay of RFB, EQFIDM will freeze in the same equilibrium background for both bouncing and inflation scenarios. In particular, for $\Omega_\chi=0.26$, we have $m_\chi\simeq 0.1~\text{keV}$, which is still not a good fermionic DM candidate as it contradicts to the condition of forming fermionic DM halos ($m_\chi\ge 0.48~\text{keV}$) \cite{Boyarsky:2008ju}.

\item Small $\langle\sigma v\rangle$. For small cross-section ($T_{fi}\gg T_r$), EQFIDM will freeze in thermal equilibrium at very early epoch that can trace a part of entropy-producing decay of RBF, so we have 
\begin{equation}\label{eq:mchifinalex}
m_\chi=\xi^{-1}\times(\Omega_\chi/2.2)\text{keV} 
\end{equation}
where $\xi\equiv \left(a_{fi}^3 s_{fi}/a_r^3s_r\right)= \left(a_{fi} T_{fi}/a_r T_r\right)^3$ is introduced with $s=\frac{2\pi^2}{45}g_\star T^3$ being the background entropy density and $g_\star$ being the background degrees of freedom. As we will show in next section, for an effective entropy-producing decay of residual bouncing fields, $\xi$ can be much smaller than $1$ and the particle mass of EQFIDM can be several orders higher than $0.1~$ keV to resolve the aforementioned tensions. 

\end{itemize}

In next section, we will focus on Small $\langle\sigma v\rangle$ case to compute $\xi$ for all possible entropy-producing decay processes of RBF.

\section{The Entropy-producing Decay of Residual Bouncing Field}
After bounce, the residual part of bouncing field such as quintom matter \cite{Cai:2009zp} starts to decay and heats the cosmic bath mildly. Without loss of generality, we take ${\Gamma_\star}$ to be the effective dissipative constant for such decay. Thus, the equations of motion governing the entropy-producing decay of RBF takes ,
\begin{equation}\label{eq:dcobfse}
\frac{d\rho_b}{dt}+3(1+w_b)H\rho_b=-{\Gamma_\star}\rho_b~,
\end{equation}
\begin{equation}\label{eq:epdeom}
\frac{d (sa^3)^{\frac{4}{3}}}{dt}=\frac{4}{3}\left(\frac{2\pi^2}{45}g_\star\right)^{\frac{1}{3}}{\Gamma_\star}\rho_b a^4~.
\end{equation}
and
\begin{equation} \label{eq:fdedepd}
H^2=\frac{1}{3M_p^2}\left(\rho_b+\rho_{\gamma}\right)
\end{equation}
where  $\rho_{\gamma}$ is the total energy density of cosmic bath, and $\rho_b$ is the energy density of RBF with $w_b\equiv p_b/\rho_b$ being its Equation of State (EoS).  Following \cite{Baltz:2001rq}, we have used $a^{-4}d(\rho_{\gamma} a^4)/dt=\Gamma_\star \rho_b$ with $s=\frac{2\pi^2}{45}g_\star T^3$ and ignored the variation of $g_\star$ during the process of interest, $dg_\star/dt=0$, to derive Eq.(\ref{eq:epdeom}). 

Integrating Eq.(\ref{eq:epdeom}) with the solution of Eq.(\ref{eq:dcobfse}), $\rho_b(t)=\rho_b(t_{fi})\left(\frac{a(t)}{a(t_{fi})}\right)^{-3(1+w_b)}e^{-{\Gamma_\star}(t-t_{fi})}$, we obtain,
\begin{equation}\label{eq:xisimplest}
\xi=\left[I+1\right]^{-\frac{3}{4}}~.
\end{equation} 
where the newly introduced dimensionless parameter takes,
\begin{equation} \label{eq:Iintgenexp}
I=\epsilon\int_{t_{fi}}^{t_r}{\Gamma_\star}e^{-{\Gamma_\star}(t-t_{fi})}\left(\frac{a(t)}{a(t_{fi})}\right)^{(1-3w_b)}dt~,
\end{equation}
with $\epsilon\equiv \rho_b(t_{fi})/\rho_{\gamma}(t_{fi})$ being the energy density ratio of RBF to cosmic bath at $T=T_{fi}$. By substituting Eq.(\ref{eq:xisimplest}) into Eq.(\ref{eq:mchifinalex}), we  obtain 
\begin{equation}
m_\chi=\left[I+1\right]^{\frac{3}{4}}\times(\Omega_\chi/2.2)\text{keV}~,
\end{equation}
which implies that, for a given value of $\Omega_\chi$, an effective post-freezing-in entropy production ($I\gg 1$) can result in a large $m_\chi$ for EQFIDM. 

Since the bouncing cosmos can be realized in many ways, the nature of RBF, $\Gamma_\star$ and $w_b$, and the value of $\epsilon$ are model-dependent. Therefore, in this paper, we are motived to perform a model-independent analysis to include all possibilities. As follows, using $\epsilon$ and $I$, the entropy-producing decay of RBF within a generic bouncing universe can be categorized into three kinds, {\bf A. Sub-dominated entropy production} ($\epsilon\ll 1$ and $I\gg 1$), {\bf B. Dominated entropy production}  ($\epsilon\gg 1$), and {\bf C. Re-dominated entropy production} ($\epsilon\ll 1$ and $I\ll 1$).
   
\subsection{Sub-domination Entropy Production ($\epsilon<1$ and $I\gg 1$)}
For sub-dominant kind, $\epsilon\ll 1$ and RBF is assumed to be always sub-dominant ($\rho_b\ll\rho_\gamma$) since $T=T_{fi}$. Thus, the entropy production takes place in a pure radiation-dominated background and the solution of Eq.(\ref{eq:fdedepd}) takes 
\begin{equation} \label{eq:cbrde}
a=a(t_{fi})\left[2H(t_{fi})(t-t_{fi})+1\right]^{\frac{1}{2}}~.
\end{equation}
By substitute it into Eq.(\ref{eq:Iintgenexp}), we obtain
\begin{equation}
I=\epsilon\int_{t_{fi}}^{t_r}{\Gamma_\star} e^{-{\Gamma_\star} (t-t_{fi})}\left[2H(t_{fi})(t-t_{fi})+1\right]^{\frac{1}{2}(1-3w_b)}dt~.
\end{equation}

As listed in Table~\ref{tab:enprotypea}, using the condition $\Gamma_\star= 2H(t_{fi})$ and $w_b=\frac{1}{3}$, this integral for sub-dominant kind can be divided into four types. In particular, for the types with large ${\Gamma_\star}$ (Type-3 and Type-4), the entropy-decaying process is so swift that the redshift can be neglected. Therefore, we obtain $I=\epsilon$, which is irrelevant to $w_b$ and $\Gamma_\star$ and find that the entropy productions in Type-3 and Type-4 are negligible as $\epsilon\ll 1$. On the other hand, for the types with small ${\Gamma_\star}$ (Type-1 and Type-2), the entropy production takes place for a longer period and the redshift should be taken into account, which leads to 
\begin{equation}
I=
\left\{  
  \begin{array} {lr}
 {\displaystyle \epsilon \times \left(\frac{2H(t_{fi})}{{\Gamma_\star}}\right)^{\frac{1}{2}(1-3w_b)}e^{\frac{{\Gamma_\star}}{2H(t_{fi})}}\Gamma\left[\frac{3}{2}(1-w_b),\frac{{\Gamma_\star}}{2H(t_{fi})} \right]\ge\epsilon~}, \quad  w_b\le \frac{1}{3} 
 \\ 
  {\displaystyle \epsilon \times \int_0^\infty \left(\frac{2H(t_{fi})}{\Gamma_\star}x+1\right)^{\frac{1}{2}(1-3w_b)}e^{-x}dx}<\epsilon~,  \qquad \qquad \qquad \quad~ w_b>\frac{1}{3}  \\
\end{array}     
\right. ,
\end{equation} 
where $\Gamma[a,x]\equiv\int_x^\infty t^{a-1}e^{-t}dt$ is the incomplete Gamma function. For Type-2, the RBF with $w_b>\frac{1}{3}$ is strongly redshifted, so that its entropy production is negligible, $I\ll \epsilon\ll 1$. For Type-1, as $w_b\le \frac{1}{3}$, its entropy production takes place longer, so that $I\ge \epsilon$. But we should note that the sub-dominant condition requires $\Gamma_\star\gg 2H(t_{fi})\left[ \epsilon^{-\frac{2}{1-3w_b}}-1\right]^{-1}$, which $I\ll 1$ (see the Appendix). It implies that the entropy production for Type-1 is also negligible. In a nutshell, we can conclude that, in the sub-domination kind, no matter what values $\Gamma_\star$ and $w_b$ are, the entropy production is always negligible. In particular, for $\Omega_\chi=0.26$, the particle mass of EQFIDM takes $m_\chi\simeq 0.1$keV in these four types.    
\begin{table*}[ht!]  
\vspace{-0.5cm}
\caption{\label{tab:enprotypea}Sub-domination Entropy Production ($\epsilon<1$ and $I\ll 1$)}
\begin{center}
\begin{tabular}{|c|c|c|}
\hline
& Small $w_b$ $\left(-1\le w_b\le \frac{1}{3}\right)$  & Large $w_b$ $\left(\frac{1}{3}\le w_b\le 1\right)$  \\
\hline
Small ${\Gamma_\star}$ & Type-1:  $\left(\Gamma_\star\gg 2H(t_{fi})\left[ \epsilon^{-\frac{2}{1-3w_b}}-1\right]^{-1}\right)$ & Type-2 \\
${\Gamma_\star}< 2H(t_{fi})$& $ \epsilon\le I\ll 1~.$&$I< \epsilon\ll 1$\\
\hline
Large ${\Gamma_\star}$ & Type-3  &Type-4\\
${\Gamma_\star}>2H(t_{fi})$& $I=\epsilon\ll 1$&$I=\epsilon\ll 1$\\
\hline
\end{tabular}
\end{center}
\vspace{-0.5cm}
\end{table*}

\subsection{Domination Entropy Production ($\epsilon>1$)}
Although cosmic bath has dominated at the end of bounce ($T=T_b$), with the expansion, RBF with a smaller $w_b$ is possible to become dominant again for a while. If EQFIDM freezes in thermal equilibrium at such RBF-dominated era, we have $\epsilon>1$ and the solution of Eq.(\ref{eq:fdedepd}) takes
\begin{equation}
a(t)=a(t_{fi})\left[\frac{3(1+w_b)}{2}\frac{2H(t_{fi})}{\Gamma_\star} \left(1-e^{-\frac{\Gamma_\star}{2}(t-t_{fi})}\right)+1\right]^{\frac{2}{3(1+w_b)}}~.
\end{equation}
By substituting it into Eq.(\ref{eq:Iintgenexp}), we obtain
\begin{equation}
I =\epsilon \int_{t_{fi}}^{t_r} {\Gamma_\star} \left[\frac{3(1+w_b)}{2}\frac{2H(t_{fi})}{\Gamma_\star}\left(1-e^{-\frac{\Gamma_\star}{2}(t-t_{fi})}\right)+1\right]^{\frac{2(1-3w_b)}{3(1+w_b)}}e^{-{\Gamma_\star} (t-t_{fi})} dt.
\end{equation}
As listed in Table~\ref{tab:enprotypeb}, using $\Gamma_\star= 2H(t_{fi})$ and $w_b=\frac{1}{3}$, this domination kind can also be categorized into four types. For Type-7 and Type-8, large ${\Gamma_\star}$ make the redshift effect negligible, which leads to $I=\epsilon$. As $\epsilon\gg 1$ in this kind, the entropy productions of Type-7 and Type-8  are effective. On the other hand, for the types with small ${\Gamma_\star}$ (Type-5 and Type-6), by taking into account the redshift effect, we obtain
\begin{equation}
I=\epsilon \frac{\left(\frac{\Gamma_\star}{2H(t_{fi})}+\frac{3(1+w_b)}{2}\right)^2}{(5-3w_b)}\left(1+\frac{3(1+w_b)}{2}\frac{2H(t_{fi})}{\Gamma_\star}\right)^{\frac{2\left(1-3w_b\right)}{3(1+w_b)}}
~.
\end{equation}
In Type-5, as $w_b\le \frac{1}{3}$, RBF are redshifted slower than cosmic bath, so its entropy production is very efficient, $I\ge \epsilon\gg 1$. But in Type-6 with $w_b> \frac{1}{3}$, RBF are redshifted faster than cosmic bath, so its entropy production is diluted, $I\le \epsilon$. In summary, the entropy production in Type-7, Type-8 and Type-5 are very effective, so that we have $m_\chi\gg 0.1~\text{keV}$ for them.  And in the Type-6, as $0\le I\le \epsilon$, $m_\chi$ ranges from $0.1~\text{keV}$ to $\epsilon^{\frac{3}{4}}\times 0.1~\text{keV}$ for different values of $w_b$ and $\Gamma_\star$.  

\begin{table*}[ht!]  
\vspace{-0.5cm}
\caption{\label{tab:enprotypeb}Dominated Entropy Production ($\epsilon\gg 1$)}
\begin{center}
\begin{tabular}{|c|c|c|}
\hline
& Small $w_b$ $\left(-1< w_b\le \frac{1}{3}\right)$  & Large $w_b$ $\left( \frac{1}{3}<  w_b\le 1\right)$  \\
\hline
Small ${\Gamma_\star}$ & Type-5 & Type-6 \\ 
${\Gamma_\star}< 2H(t_{fi})$& $I=\epsilon \frac{\left(\frac{\Gamma_\star}{2H(t_{fi})}+\frac{3(1+w_b)}{2}\right)^2}{(5-3w_b)}\left(1+\frac{3(1+w_b)}{2}\frac{2H(t_{fi})}{\Gamma_\star}\right)^{\frac{2\left(1-3w_b\right)}{3(1+w_b)}}
$&$I=\epsilon \frac{\left(\frac{\Gamma_\star}{2H(t_{fi})}+\frac{3(1+w_b)}{2}\right)^2}{(5-3w_b)}\left(1+\frac{3(1+w_b)}{2}\frac{2H(t_{fi})}{\Gamma_\star}\right)^{\frac{2\left(1-3w_b\right)}{3(1+w_b)}}$\\
& $\ge \epsilon\gg 1$ & $< \epsilon$\\
\hline
Large ${\Gamma_\star}$ & Type-7  &Type-8\\
${\Gamma_\star}>2H(t_{fi})$& $I=\epsilon\gg 1$&$I=\epsilon\gg 1$\\
\hline
\end{tabular}
\end{center}
\vspace{-0.5cm}
\end{table*}

\subsection{Re-domination Entropy Production ($\epsilon <1$ and $I\gg 1$)}

In re-domination kind, we also have $\epsilon\ll 1$ and ${\Gamma_\star}< 2H(t_{fi})$, which are the same to Type-1 in the sub-dominant kind. But now we investigate the other part of parameter region, $\left(\Gamma_\star\ll 2H(t_{fi})\left[ \epsilon^{-\frac{2}{1-3w_b}}-1\right]^{-1}\right)$, which allows RBF to become re-dominant after EQFIDM freezes in. To facilitate our analysis, we divide the history for this kind into three phases,
\begin{enumerate}
\item Pre-re-domination phase ($t_{fi}\le t\le t_{rd}$). RBF is sub-dominated, so that the entropy production is negligible;    
\item Re-domination phase ($t_{rd}\le t\le t_{rsd}$).  RBF becomes re-dominant and results in an effective entropy production;  
\item Post-re-dominated phase ($t>t_{rsd}$). As RBF is depleted, cosmic bath becomes dominant again, so that the entropy production is also negligible, 
\end{enumerate}
where $t_{rd} =\left[2H(t_{fi})\right]^{-1} \left(\epsilon^{-\frac{2}{1-3w_b}}-1\right)$ can be obtained from $\rho_\gamma(t_{rd})=\rho_b(t_{rd})$ and $t_{rsd}\rightarrow \infty$ can be taken for simplicity. 

As the entropy production is mainly contributed from the Re-domination phase, we have
\begin{equation}\label{eq:itype9a}
I=\epsilon \frac{\rho_b(t_{rd})}{\rho_b(t_{fi})} \left(\frac{a(t_{rd})}{a(t_{fi})}\right)^4\int_{t_{rd}}^{t_{rsd}}{\Gamma_\star}\frac{\rho_b}{\rho_b(t_{rd})} \left(\frac{a}{a(t_{rd})}\right)^4dt~.
\end{equation}
By integrating it with the background solution,
\begin{equation} 
a(t)=a(t_{rd})\left[\frac{3(1+w_b)}{2}\frac{2H(t_{fi})}{\Gamma_\star}\left(1-e^{-\frac{\Gamma_\star}{2}(t-t_{fi})}\right)+1\right]^{\frac{2}{3(1+w_b)}}~,
\end{equation} 
and using $ \frac{\rho_b(t_{rd})}{\rho_b(t_{fi})} \left(\frac{a(t_{rd})}{a(t_{fi})}\right)^4=\epsilon^{-1}$ (see Appendix), we obtain
\begin{equation}
I=\frac{\left(\frac{\Gamma_\star}{2H(t_{fi})}+\frac{3(1+w_b)}{2}\right)^2}{(5-3w_b)}\left(1+\frac{3(1+w_b)}{2}\frac{2H(t_{fi})}{\Gamma_\star}\right)^{\frac{2\left(1-3w_b\right)}{3(1+w_b)}}~,
\end{equation} 
which is irrelevant to $\epsilon$ (Note that, this result is different from Type-5 by a factor $\epsilon$). Notably, this result implies that, if RBF can become re-dominant, its entropy production solely depends on its fundamental nature, $\Gamma_\star/2H(t_{fi})$ and $w_b$. This is because the entropy production is mainly contributed during the Re-domination phase ($t_{rd}\le t\le t_{rsd}$), which duration and other details only depends on $\Gamma_\star/2H(t_{fi})$ and $w_b$. In other words, if $\Gamma_\star$ is small enough, $\Gamma_\star\ll 2H(t_{fi})\left[ \epsilon^{-\frac{2}{1-3w_b}}-1\right]^{-1}$, the decay of RBF can also lead to a significant contribution to the post-freezing-in entropy production even if RBF only contributes a very small portion to total energy density at freezing-in. 

In Table~\ref{tab:enprotypec}, we list Type-9. As we will show, for different $\Gamma_\star/2H(t_{fi})$ and $w_b$, the particle mass of EQFIDM in this type can range from sub-keV scale to super-keV scale.   
\begin{table*}[ht!]  
\vspace{-0.5cm}
\caption{\label{tab:enprotypec}Re-dominated Entropy Production ($\epsilon\gg 1$)}
\begin{center}
\begin{tabular}{|c|c|c|}
\hline
& Small $w_b$ $\left(-1< w_b\le \frac{1}{3}\right)$  & Large $w_b$ $\left( \frac{1}{3}\le  w_b\le 1\right)$  \\
\hline
Small ${\Gamma_\star}$ & Type-9:  $\left(\Gamma_\star\ll 2H(t_{fi})\left[ \epsilon^{-\frac{2}{1-3w_b}}-1\right]^{-1}\right)$& --- \\
${\Gamma_\star}< 2H(t_{fi})$& $I=\frac{\left(\frac{\Gamma_\star}{2H(t_{fi})}+\frac{3(1+w_b)}{2}\right)^2}{(5-3w_b)}\left(1+\frac{3(1+w_b)}{2}\frac{2H(t_{fi})}{\Gamma_\star}\right)^{\frac{2\left(1-3w_b\right)}{3(1+w_b)}}~.$&\\
\hline
Large ${\Gamma_\star}$ & ---  & ---\\
\hline
\end{tabular}
\end{center}
\vspace{-0.5cm}
\end{table*}

To sum up, by performing a model-independent analysis of the entropy-producing decay of RBF within a generic bouncing universe, we have obtained all three kind of entropy production, which includes nine different types. Specifically, the entropy production in Type-5, Type-7 and Type-8 are very effective, which leads $m_\chi\gg 0.1~\text{keV}$. In  Type-1, Type-2, Type-3 and Type-4, the entropy productions are inefficient, so $m_\chi\simeq 0.1~\text{keV}$. In Type-6, as the entropy production can be diluted by redshift, $0.1~\text{keV}\le m_\chi\ll \epsilon^{\frac{3}{4}}\times 0.1~\text{keV}$. And in the most interesting case, Type-9, its entropy production is solely determined by the fundamental nature of RBF,  $\Gamma_\star/2H(t_{fi})$ and $w_b$, rather than $\epsilon$. With different values of $\Gamma_\star/2H(t_{fi})$ and $w_b$, the particle mass of EQFIDM  can range from sub-keV scale to super-TeV scale. In next section, we are investigate how current astrophysical observation can constrain these types.

\section{Astrophysical Implications} 

As collisionless particle, EQFIDM's motion can smooth the inhomogeneities of spacetime and matter at late-time evolution and induce a suppression of linear matter spectrum at a scale comparable to its free-streaming length \cite{Zentner:2003yd},
\begin{equation}
L_{fs}\simeq 0.11\left(\frac{\Omega_\chi h^2}{0.15}\right)^{\frac{1}{3}}\left(\frac{m_\chi}{\text{keV}}\right)^{-\frac{4}{3}} \text{Mpc}=1.8\left(I+1\right)^{-1} \text{Mpc}~,
\end{equation} 
where Eq.(\ref{eq:mchifinalex}) and Eq.(\ref{eq:xisimplest}) have been used at the last step. According to this equation, with an inefficient entropy production ($I\le 1$), the suppression takes place around $L_{\text{fs}}\simeq 1.8~\text{Mpc}$, which is too large to be compatible with current Lyman-$\alpha$ forest observation on linear matter spectrum -- as we will show. Only if the entropy production from RBF is effective ($I\gg 1$), the suppression scale can be as small as desired.  

To explicitly plot the linear matter spectrum in presence of EQFIDM, we adopt the numerical simulation results presented in Ref.\cite{Markovic:2013iza}. Due to the free-streaming of EQFIDM, the linear matter spectrum takes 
\begin{equation}\label{eq:lmseqfiDM}
P(k)=P_{cdm}(k)\mathcal{T}^2(k)~,
\end{equation} 
where $P_{cdm}(k)$ is the spectrum for the extreme cold dark matter (CDM) candidate and $\mathcal{T}=\left[1+(\alpha k)^{2\nu}\right]^{-\frac{5}{\nu}}$ is the transfer function with $\nu=1.12$ and $\alpha=0.57\left(\frac{\Omega_\chi}{0.26}\right)\left(I+1\right)^{-0.83}$.

\begin{figure}[htp!]
\centering
\includegraphics[width=0.8\textwidth]{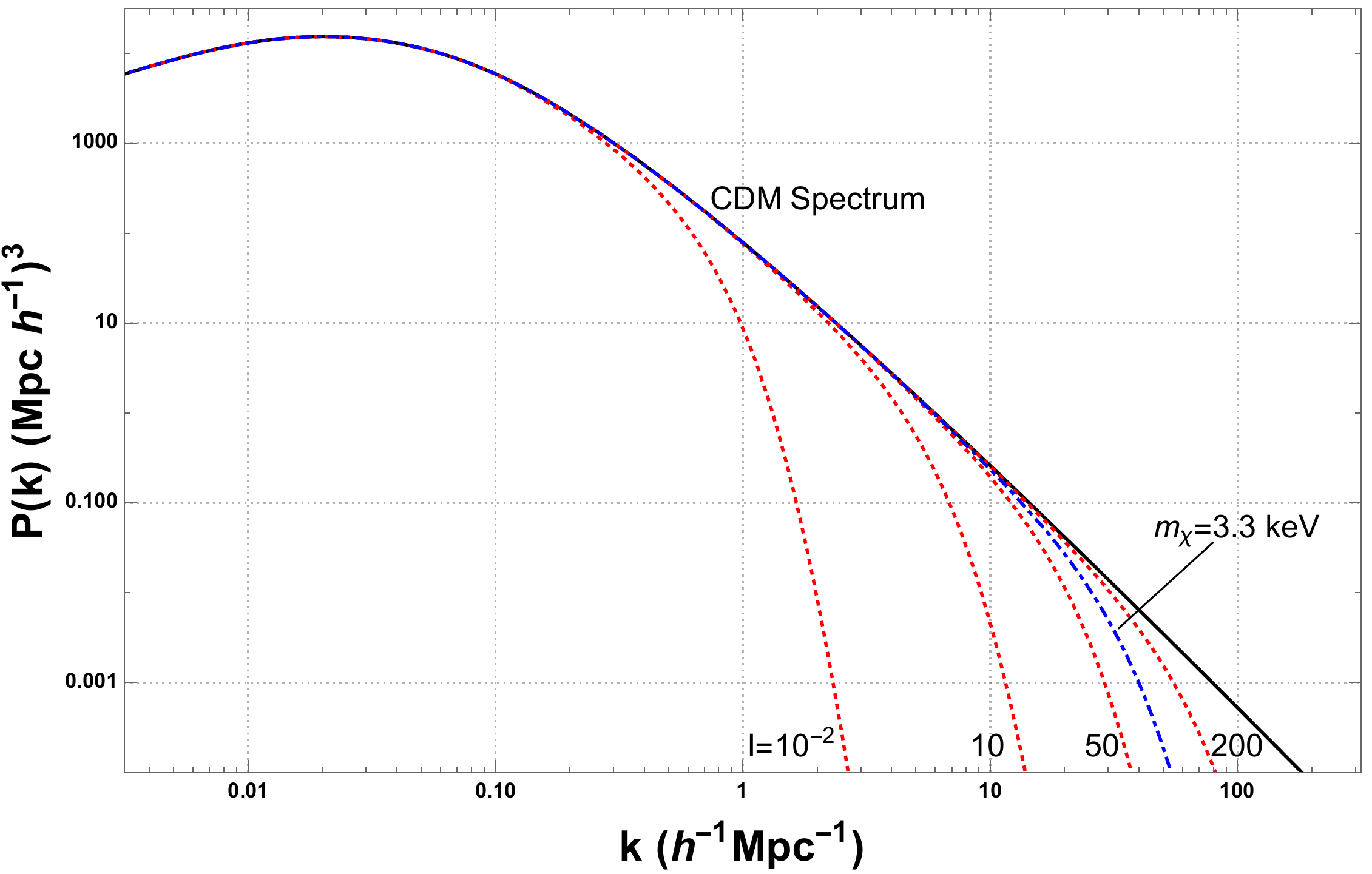}
\caption{The power spectrum of linear matter perturbation for EQFIDM with $I=\{10^{-2}, 10, 50, 200\}$ (the dotted lines from left to right), warm DM candidate with $m_\chi=3.3~\text{keV}$ (the dashed line), and fiducial CDM candidate (the solid line). }
\label{fig:WDMpower.pdf}
\end{figure}

In Fig.\ref{fig:WDMpower.pdf}, we plot the linear matter spectrum for EQFIDM with $I=\{10^{-2}, 10, 50, 200\}$ (the dotted lines from left to right), warm DM candidate with $m_\chi=3.3~\text{keV}$ (the dashed line), and fiducial CDM candidate (the solid line). According to current Lyman-$\alpha$ forest observation on linear matter spectrum, DM candidate, which particle mass is smaller than $3.3~\text{keV}$, has been excluded as its free-streaming length (the suppression scale) is too large. Therefore, as shown in Fig.\ref{fig:WDMpower.pdf}, EQFIDM candidates in the types with negligible entropy production ($I\ll 1$: Type-1, Type-2, Type-3 and Type-4) have been excluded. But, for the types with effective entropy production ($I\gg 10^2$: Type-5, Type-7, Type-8 ), EQFIDM candidates is cold enough that compatible with current observation. For Type-6 ($0\le I\ll \epsilon$, $\epsilon\gg 1$), it requires $\epsilon\gg I\gg 10^2$ to be compatible with current observation. And for the most interesting case, Type-9, the particle mass of EQFIDM is irrelevant to $\epsilon$. As we will show, it can also be compatible with current observation in a sizeable parameter space of $\Gamma_\star/2H(t_{fi})$ and $w_b$.

\begin{figure}[htp!]
\centering
\includegraphics[width=0.8\textwidth]{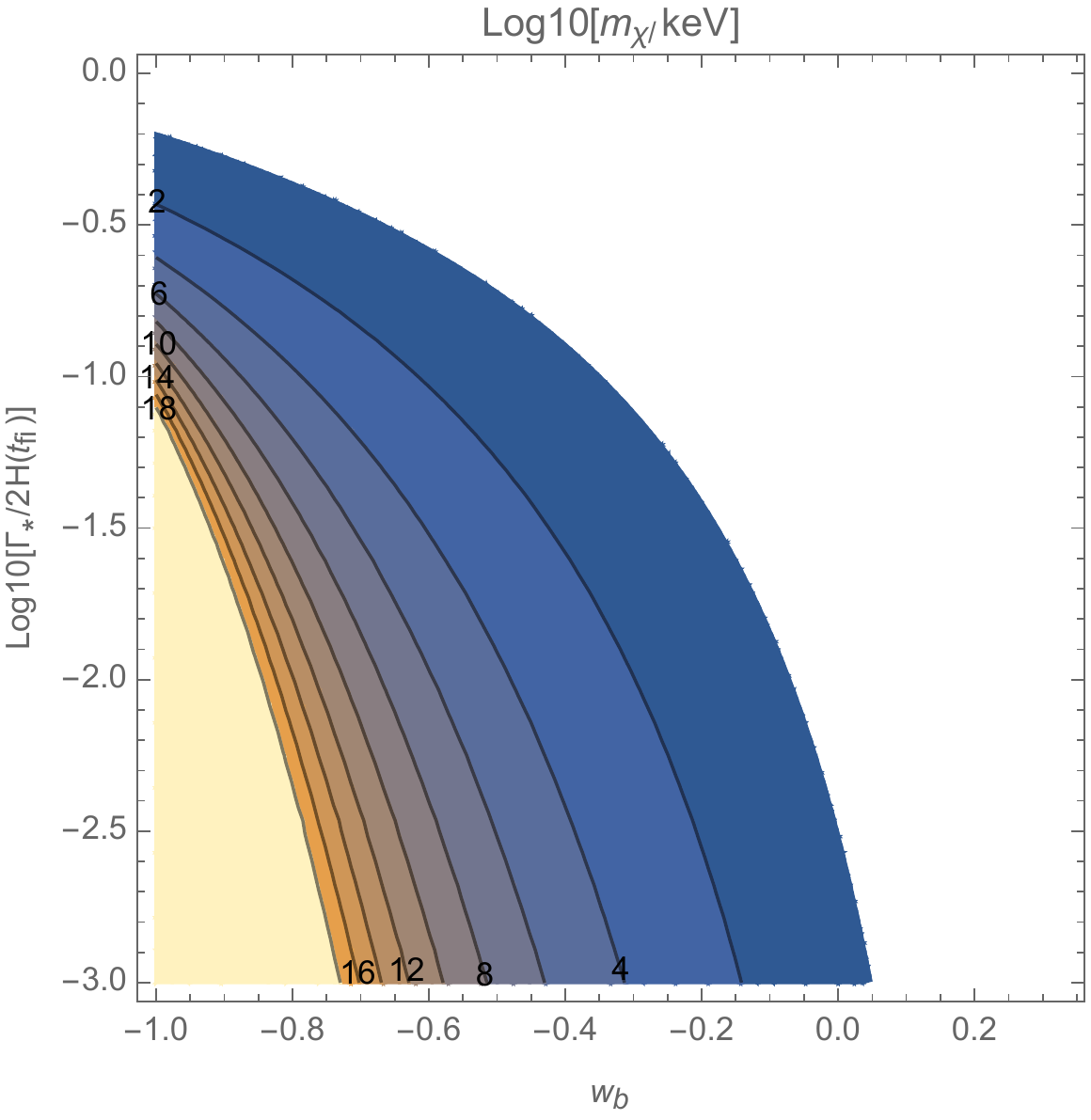}
\caption{The contour plot for the particle mass of EQFIDM in Type-9 with respect to the fundamental nature of RBF, $w_b$ and $\Gamma_\star/2H(t_{fi})$. The blank region has been excluded by current Lyman-$\alpha$ forest observation.}
\label{fig:CPparticlemass.pdf}
\end{figure}

In Fig.\ref{fig:CPparticlemass.pdf}, we plot the exclusion region ($m_\chi\le 3.3~\text{keV}$) for Type-9 with respect to $w_b$ and $\Gamma_\star/2H(t_{fi})$. As it shows, there is a sizeable region that EQFIDM can be compatible with current Lyman-$\alpha$ forest observation on linear matter spectrum. Specifically, in the the outermost region ($3.3~\text{keV}\le m_\chi\le 100~\text{keV}$), EQFIDM is warm as the entropy production from RBF is mild. But in the inner regions, with very effective entropy productions, its particle mass is very large and even can exceeds $\text{TeV}$ scale. In these regions, EQFIDM is belong to cold DM candidate, which is worthy for further study. 

Furthermore, according to the above analysis, both Type-9 (for the outermost region) and Type-6 (with large $\epsilon$) can serve as warm DM candidates. Therefore, we are motivated to investigate their potential on alleviating the well-known small-scale crisis (SSC). As one of the most outstanding puzzles in modern astrophysics, the small-scale crisis refers to an underlying discrepancy between cosmological prediction and astrophysical observation, which mainly includes three aspects, {\it the missing satellites problem}, {\it the cusp v.s. core problem}, and {\it the too-big-to fail problem} \cite{Weinberg:2013aya}. More specifically, the current observations on the sub-galactic scale indicates that less satellites are observed than the prediction from N-body numerical simulation based on the standard $\Lambda$CDM cosmology. In additional, the observed satellites are also less concentrated than the theoretical prediction. Although such crisis may be relieved by further elucidation on the subtle baryonic physics and/or better interpretation of the observational data \cite{Simon:2007dq, Diemand:2008in, Ackermann:2011wa, Amendola:2012ys, Koposov:2015cua, Drlica-Wagner:2015xua, Kim:2017iwr}, it may also implies a suppression on the matter density perturbation at small scale ($10~\text{kpc}\le l\le 200~\text{kpc} $) \cite{Markovic:2013iza}. To realize such suppression on matter density perturbation, many salient mechanisms, which are beyond the simple inflation and cold DM, have been proposed, mainly including the non-simple-inflation-inspired category such as broken-scale-invariance inflation \cite{Kamionkowski:1999vp, Yokoyama:2000tz, Kobayashi:2010pz, Nakama:2017ohe} and the non-simple-DM-inspired category such in \cite{Lin:2000qq, Sigurdson:2003vy, Kusenko:2009up, Rocha:2012jg, Hochberg:2014dra, Foot:2014uba, Hochberg:2014kqa, Tulin:2017ara}. Among them, the warm dark matter candidates can also induce such desired suppression as its motion can smooth the anisotropies and inhomogeneities at a scale comparable with its free-streaming length \cite{Bode:2000gq,Colin:2000dn, Viel:2011bk, Maccio:2012rjx, Viel:2013fqw,  Murgia:2017lwo}. 

\begin{figure}[htp!]
\centering
\includegraphics[width=0.8\textwidth]{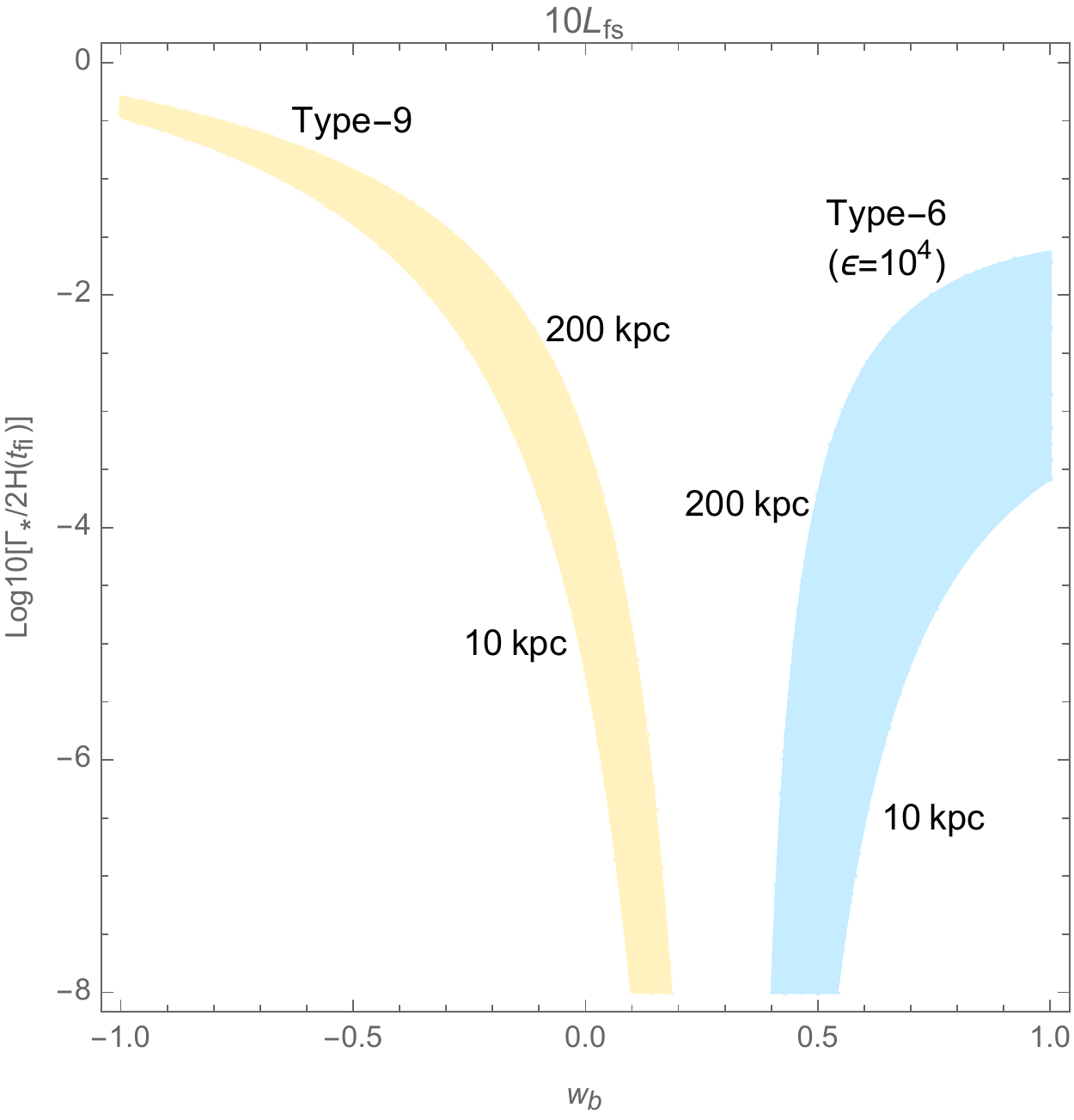}
\caption{The parameter region of Type-9 and Type-6 (with $\epsilon=10^4$) for $3.44~\text{keV}\le m_\chi\le 32.6~\text{keV}$, which can alleviate SSC by inducing a suppression of linear matter spectrum at $10~\text{kpc}\le 10 L_{\text{fs}}\le 200~\text{kpc}$.}
\label{fig:SSCfig.pdf}
\end{figure}

In Fig.\ref{fig:SSCfig.pdf}, we illustrate the parameter region of Type-9 and Type-6 (with $\epsilon=10^4$) for $3.44~\text{keV}\le m_\chi\le 32.6~\text{keV}$, which can alleviate SSC by inducing a suppression of linear matter spectrum at $10~\text{kpc}\le 10 L_{\text{fs}}\le 200~\text{kpc}$,  in $(\Gamma_\star/2H(t_{fi}), w_b)$ space. As it shows, both types have sizeable parameter regions that can alleviate SSC. Especially, the result from Type-9 implies that, even though RBF may only contribute a very small portion of total energy density at freezing-in and have a very tiny decay rate, it also can alleviate SSC. Except these two, the rest seven types can not alleviate the SSC as the entropy productions are either too effective or too inefficient. 

\section{Summary}
In this paper, by carefully examining thermal decoupling condition within a generic bouncing universe, we unveil a new fermionic DM candidate (EQFIDM), which can freeze in thermal equilibrium shortly after bounce. As the relic abundance of EQFIDM depends on the post-freezing-in entropy production, we use it as a novel probe to trace the entropy production from residual bouncing field (RBF). Specifically, we present a model-independent formalism to obtain all nine possible types of entropy-producing decay process of RBF within the generic bouncing cosmos (Type-1 -- Type-9).  We find that, due to the entropy productions in these types, the particle mass of EQFIDM can range from sub-keV to super-TeV. 

Using current Lyman-$\alpha$ forest observation on linear matter spectrum, we find that, although four types (Type-1 -- Type-4) have been excluded for their tiny entropy productions, the rest five types, which suggest warm and cold EQFIDM candidates with effective entropy productions, are accommodated with current observation in the whole (Type-5, Type-7 \& Type-8) or a part of parameter region (Type-6 \& Type-9). In particular, in the re-dominant type (Type-9), the entropy production is irrelevant to the energy density ratio at freezing-in, $\epsilon$, which implies that the particle mass of EQFIDM in this type is solely determined by the fundamental nature of RBF ($\Gamma_\star/2H(t_fi)$ and $w_b$). Then we illustrate how its nature can be constrained by the current observation. 

Furthermore, as both Type-9 and Type-6 are allowed to serve warm DM candidates within a part of their parameter regions, we also investigate their potential on alleviating the small-scale crisis (SSC). It turns out both of them have sizeable parameter region that can alleviate the SSC. Especially, the result from Type-9 indicates, even though RBF may only contribute a very small portion of total energy density at freezing-in and have a very tiny decay rate, it also can alleviate SSC. 

To sum up, our model-independent analysis may shed light on model building in bouncing cosmology. Optimized by future astrophysical observations, the entropy-producing decay processes of RBF and the properties of EQFIDM in different types may be further constrained. And, hopefully, the result from Type-9  may help us better understand SSC with respect the  post-bounce entropy production. At the end, we wish to highlight that a similar analysis can also be applied to NEQDM candidate as its abundance is also sensitive to the entropy-producing decay of RBF.

\section{Acknowledgments} 
I thank Yeuk-kwan Edna Cheung for useful discussion. This work has been supported in parts by the National Natural Science Foundation of China (11603018, 11963005, 11775110, 11433004, 11690030), and Yunnan Provincial Foundation (No.2016FD006, 2015HA022, 2015HA030, 2019FY003005).

\section{Appendix: Re-domination condition}
To ensure the RBF is alway sub-dominated for Type-1, we have $\rho_\gamma(t_{fi}+\Delta t)\gg \rho_b(t_{fi}+\Delta t)$ with $\Delta t=\Gamma_\star^{-1}$ being the half-life of RBF, which leads 
\begin{equation}
\left(\frac{a(t_{fi})}{a(t_{fi}+\Gamma_\star^{-1})}\right)^{1-3w_b}\gg \epsilon~,
\end{equation}
where $\epsilon\equiv \rho_b(t_{fi})/\rho_\gamma(t_{fi})\ll 1$ for the Sub-dominant kind and Re-dominant kind. By substituting Eq.(\ref{eq:cbrde}) into it, we obtain
\begin{equation}
\Gamma_\star\gg 2H(t_{fi})\left[ \epsilon^{-\frac{2}{1-3w_b}}-1\right]^{-1}~,
\end{equation}
which is the sub-dominant condition for RBF. Additionally, by applying it to Type-I ($\Gamma_\star<2H(t_{fi})$, $-1\le w_b\le \frac{1}{3}$ and $\epsilon<1$), we can obtain
\begin{equation}
I\ll \epsilon\times \epsilon^{-1} e^{\frac{{\Gamma_\star}}{2H(t_{fi})}}\Gamma\left[\frac{3}{2}(1-w_b),\frac{{\Gamma_\star}}{2H(t_{fi})} \right]\simeq \Gamma\left[\frac{3}{2}(1-w_b)\right]\simeq 1~,
\end{equation}
which indicates that the entropy production in Type-I is also negligible, where $\Gamma[n]$ is the Gamma function. 

Accordingly, we can obtain the re-dominant condition for Type-9,
\begin{equation}
\Gamma_\star\ll 2H(t_{fi})\left[ \epsilon^{-\frac{2}{1-3w_b}}-1\right]^{-1}~,
\end{equation}
which straightforwardly leads to a useful relation for Type-9,
\begin{equation}
\frac{\rho_b(t_{rd})}{\rho_b(t_{fi})} \left(\frac{a(t_{rd})}{a(t_{fi})}\right)^4=e^{-\frac{\Gamma_\star}{2H(t_fi)}\left(\epsilon^{-\frac{2}{1-3w_b}}-1\right)} \epsilon^{-\frac{2}{1-3w_b}\cdot \frac{1-3w_b}{2}} =\epsilon^{-1}~.
\end{equation}

\end{document}